# Impurity effect on weak anti-localization in topological insulator Bi$_2$Te$_3$

Hong-Tao He[1], Gan Wang[1,2], Tao Zhang[1,2], Iam-Keong Sou[1], George K. L. Wong[1], and Jian-Nong Wang[1,*]

[1]*Department of Physics, The Hong Kong University of Science and Technology, Clear Water Bay, Hong Kong, China*

[2]*Nano Science and Technology Program, The Hong Kong University of Science and Technology, Clear Water Bay, Hong Kong, China*

Hai-Zhou Lu[3], Shun-Qing Shen[3], and Fu-Chun Zhang[3]

[3]*Department of Physics, The University of Hong Kong, Hong Kong, China*

Abstract

We study weak anti-localization (WAL) effect in topological insulator Bi$_2$Te$_3$ thin films at low temperatures. Two-dimensional WAL effect associated with surface carriers is revealed in the tilted magnetic field dependence of magneto-conductance. Our data demonstrates that the observed WAL is robust against deposition of non-magnetic Au impurities on the surface of the thin films. But it is quenched by deposition of magnetic Fe impurities which destroy the $\pi$ Berry's phase of the topological surface states. The magneto-conductance data of a 5 nm Bi$_2$Te$_3$ film suggests that a crossover from symplectic to unitary classes is observed with the deposition of Fe impurities.

---

[*] Corresponding author email address: phjwang@ust.hk

Three-dimensional (3D) topological insulators (TIs) are band insulators with gapless, Dirac-particle like surface states (SSs) that are protected by time-reversal symmetry (TRS) [1, 2]. The topological SS has a number of interesting properties [3-6] that result from the fact that the electron spin is locked with its momentum. 3D TI was first observed in $Bi_{1-x}Sb_x$ alloys [7]. 3D TIs with a single Dirac cone on the surface was first proposed theoretically and then observed experimentally in $Bi_2Se_3$ and $Bi_2Te_3$ [8-10]. Recent transport studies of $Bi_2Se_3$ and $Bi_2Te_3$ have revealed various quantum interference phenomena, indicating the existence of topological SSs even after the sample being exposed in air and subjected to microelectronic processing [11-15]. However, transport experiments on 3D TI have been hindered by residual bulk carriers since the Fermi levels of as-grown $Bi_2Se_3$ or $Bi_2Te_3$ samples are usually located either in the conduction band [10] or the valence band [16]. To study surface transport properties, one has to shift the Fermi level into the band gap by Ca dopants [11] or increase surface-to-volume ratio [12]. In this Letter, we report the investigation of weak anti-localization (WAL) effect in n-type $Bi_2Te_3$ thin films at low temperatures ($T$) and in titled magnetic fields as well as the effect of the impurities on the WAL. The titled magnetic field measurement allows us to distinguish 2D WAL effect associated with the topological SSs from 3D bulk effect. We show that magnetic Fe impurities deposited on the surface of $Bi_2Te_3$ films suppress 2D WAL effect while non-magnetic Au impurity deposition has little effect on 2D WAL. This reveals that the observed 2D WAL effect mainly results from the top topological SS and that the $\pi$ Berry's phase of the topological SS is destroyed by magnetic impurities but robust against non-magnetic impurities. A crossover from symplectic to unitary classes is also clearly observed in the magneto-conductance of 5 nm $Bi_2Te_3$ films when magnetic Fe impurities are deposited on the film surface.

Our $Bi_2Te_3$ films were grown along c-axis by molecular beam epitaxy (MBE) on (111) semi-insulating GaAs substrates on which an undoped ZnSe buffer layer has been deposited. To prevent the $Bi_2Te_3$ film surface from contaminations, the films were capped by a 3 nm undoped ZnS top layer. The magneto-transport data obtained from samples with and without ZnS capping showed little difference. The as-grown films (5 nm and 50 nm) are metallic as indicated by their temperature-dependence of resistivity (see Fig. 1 (a)). The electron concentration obtained from the Hall effect data is 0.8 – 1.9 x $10^{20}$ cm$^{-3}$ for 5 nm samples and 1.3 – 1.6 x $10^{19}$ cm$^{-3}$ for 50 nm samples. A schematic band structure with gapless Dirac cone is illustrated in Fig. 1 (a) with Fermi level $E_F$ in the conduction band. Note that a hybridization gap due to inter-surface coupling has been observed in 5 nm $Bi_2Se_3$ films [17], but not clearly in 5 nm $Bi_2Te_3$ films [18]. To facilitate magneto-conductance measurements the $Bi_2Te_3$ films were fabricated into Hall bars of dimension 200 μm x 100 μm (*l* x *w*, see Fig. 1 (a)) using standard photolithographic processes with Cr(10nm)/Au(150nm) metal ohmic contacts deposited by thermal evaporation. Four-probe magneto-resistance (MR) was measured in a 14 T Quantum Design PPMS system which has a base temperature of 2 K. In MR measurements, we fixed the $B$ field in the $z$ direction and the Hall voltage probes in the $y$ direction as shown in Fig. 1 (a). In our experimental system the Hall bar can be rotated about the $y$ axis using a rotational sample holder. In Fig. 1 (b), we show the MR data of a 5 nm $Bi_2Te_3$ film at low $T$ with $B$ field applied perpendicular to the film plane, *i.e.* $\theta = 90°$ where $\theta$ is the angle between $B$ and the current $I$. A sharp resistance dip is clearly observed at $T$ = 2K, indicating the presence of WAL effect [19]. As $T$ increases, the MR dip at low $B$ is broadened and finally disappears due to the decrease of the phase coherent length at higher $T$.

This WAL effect was investigated in tilted $B$ fields at $T$ = 2K, as shown in Fig. 2 (a) for a 5 nm film. At $\theta = 0°$, i.e. with $B$ in the film plane, the MR dip feature disappears completely and the MR shows a parabolic B-field dependence. This semi-classical $B^2$ dependence results from the Lorentz deflection of carriers. According to the Kohler's rule [20], $R(B)/R(B=0) \approx 1 + (\mu B)^2$, the film mobility ($\mu$) is estimated to be 521 $cm^2V^{-1}s^{-1}$. By subtracting this background parabola from magneto-conductance, we obtain the WAL-induced quantum corrections to magneto-conductance, $\Delta G(\theta, B) = 1/R(\theta, B) - 1/R(0, B)$. Fig. 2 (b) shows $\Delta G$ as a function of the normal component of B-field, *i.e.* $B \sin\theta$, for various tilting angles. It can be clearly seen that all $\Delta G(\theta, B)$ curves coincide with each other at low magnetic fields. This behavior indicates that the observed WAL effect results from the orbital motion of carriers since the orbital term depends only on the normal component of the magnetic field. In higher $B$ fields, suppression of the WAL effect by the spin term is observed. This spin effect arises from the Zeeman splitting and it depends on the magnitude of the $B$ field [21]. The tilted $B$ field data clearly show that the observed WAL in 5 nm $Bi_2Te_3$ film is of 2D nature. It could result from the thin film itself (5 nm thin film is a quasi-2D system) due to the strong spin-orbit coupling in $Bi_2Te_3$, from the 2D topologically protected SSs, or from both. To further clarify the origin of this WAL effect, we investigated the MR properties of a 50 nm $Bi_2Te_3$ film, which is a 3D system. We expected this thicker film should exhibit different MR properties from the 5 nm thin film [17].

Fig. 2 (c) shows the MR data obtained in tilted $B$ fields at $T$ = 2K for a 50 nm $Bi_2Te_3$ thin film. Similar to the 5 nm sample, the WAL effect depends on the tilt angle. It is most pronounced at $\theta = 90°$, and becomes weaker with decreasing $\theta$. In sharp contrast to the 5 nm film, the WAL in the 50 nm sample can be observed even when the $B$ field is in-plane ($\theta$=0), as we can see from Fig. 2 (c). We ascribe the WAL effect observed at $\theta$=0 to 3D WAL effect that has its origin in the bulk of the 50 nm Bi$_2$Te$_3$ film. Since 3D WAL effect does not depend on the tilt angle of the B-field [22], we can subtract 3D WAL contribution from the magneto-conductance data obtained at other angles, i.e. $\Delta G(\theta, B) = 1/R(\theta, B) - 1/R(0, B)$. In Fig. 2 (d), we plot $\Delta G$ as a function of the normal component of $B$. The $\Delta G$ curves in Fig. 2 (d) coincide with each other in low $B$ fields but deviate from each other in high $B$ fields, similar to the data obtained from the 5 nm film. After subtracting out 3D bulk WAL effect, we obtain the WAL effect due to topologically protected 2D SSs of the Bi$_2$Te$_3$ film. Previous spin-resolved photoemission study performed on Bi$_2$Te$_3$ has revealed that the Fermi surface of topological SS exhibits unconventional quantum spin textures [4]. A surface electron moving in circles above the Dirac point could acquire a $\pi$ Berry's phase [23] while the electron spin is rotated by $2\pi$. This $\pi$ Berry's phase changes the interference of a pair of time reversed paths from being constructive to destructive. As a result, topological SSs are expected to exhibit 2D WAL effect, in agreement with our observations shown in Fig. 2 (d).

According to 2D localization theory [19, 24, 25], 2D magneto-conductivity $\sigma_{2D}$ is described by the following equation under the assumption that the inelastic scattering time is much longer than both the elastic and spin-orbit scattering times:

$$\Delta\sigma_{2D} = \sigma_{2D}(B) - \sigma_{2D}(0) = -\frac{\alpha e^2}{2\pi^2\hbar}\left[\ln\frac{\hbar}{4Bel_\varphi^2} - \psi\left(\frac{1}{2} + \frac{\hbar}{4Bel_\varphi^2}\right)\right] \quad (1)$$

where $e$ is the electronic charge, $\hbar$ is the Plank's constant, $l_\varphi$ is the phase coherent length, and $\psi(x)$ is the digamma function. $\alpha$ should be equal to 1, 0, and -1/2 for the orthogonal, unitary, and symplectic cases, respectively [19, 26]. Since strong spin-orbit interaction is expected in topological SSs, the assumption used in deducing Eq. (1) should be valid. The magneto-conductance tip at zero-field of the $\Delta G(\theta = 90, B)$ curve can be well fitted with this equation, as shown by the solid fitting curve in Fig. 2 (d). The fitting yields $\alpha$ =-0.39 and $l_\varphi$ =331 nm, respectively. Quantum spin texture systems, such as the 2D SS of a 3D TI, belong to the symplectic class and $\alpha$ should be equal to -1/2 for one topological SS and -1 in a film with one top and one bottom SS. The value of the fitting parameter $\alpha$ obtained from the fit is -0.39, indicating that there is only one SS contributing to the 2D WAL effect shown in Fig. 2 (d). This is in agreement with a recent study of WAL effect in Bi$_2$Se$_3$ film with a back gate [14]. The obtained phase coherent length $l_\varphi$ =331 nm is comparable to $l_\varphi$

~ 500 nm estimated from the Aharonov-Bohm interference study on $Bi_2Se_3$ nano-ribbons [12]. In contrast to the recent observations of localization phenomena in graphene [27, 28], the competition between localization and anti-localization due to short-range inter-valley scatterings [24, 29] is absent in our samples because in our films only one Dirac cone dominates the transport properties.

Spin-momentum locking of Dirac fermions leads to the $\pi$ Berry's phase of SSs after the electron has undergone an adiabatic cycling in momentum space. This Berry phase is thus vulnerable to spin-dependent interactions that break the TRS, *e.g.* the exchange interactions with magnetic impurities deposited on the surface. To further confirm the observed 2D WAL effect is indeed caused by the $\pi$ Berry's phase, we have grown $Bi_2Te_3$ thin films containing magnetic Fe impurities deposited at the interface between the ZnS capping layer and the $Bi_2Te_3$ active layer. Samples with Fe deposition of 1 monolayer (ML) and 0.3 ML (as estimated using the MBE growth rate) were grown. For comparison, a 5 nm $Bi_2Te_3$ thin film with one ML non-magnetic Au-deposition at the interface was also prepared and studied. Data obtained from samples with Fe (or Au) deposited on the surface show that the resistivity of these films is similar to that of pristine $Bi_2Te_3$ thin films, indicating that Fe or Au atoms deposited at the interface can be reasonably assumed to form well separated clusters and hence they do not contribute directly to the conductance of our films. It is also possible that the deposited Fe atoms form magnetic Fe-Te complexes on the film surface, but these complexes should also impose similar magnetic perturbation to the topological SSs just like in the case of Fe clusters [30]. Although Fe atoms deposited on $Bi_2Se_3$ surfaces are electron donors [30], the observed electron density actually decreases a little in our $Bi_2Te_3$ films with Fe deposition (from 1.35 to 1.02 $\times 10^{20}$ cm$^{-3}$ and from 1.6 to 1.34 $\times 10^{19}$ cm$^{-3}$ for 5 and 50 nm films respectively). Different growth techniques used by us might be the cause of this difference. Additionally, no anomalous Hall effect is observed in Fe-deposited samples, which seems to suggest that the magnetizations of the Fe clusters are random in direction. The exchange interaction between the surface states and the random Fe impurity spins can result in a random spatial distribution of local gap $\Delta$ in the gapless spectrum of the surface states since TRS is broken locally by the Fe clusters. As a result, suppression of 2D WAL effect is expected to occur with Fe deposition. The MR of thin films with Fe deposition was then measured in tilted $B$ fields at $T = 2$ K. The obtained magneto-conductance, $\Delta G(\theta=90,B) = 1/R(\theta=90,B) - 1/R(0=90,B)$, is plotted as a function of the normal $B$ component and shown in Fig. 3 (red open squares for 1 ML and blue open squares for 0.3 ML Fe-deposition, respectively). The same quantity, $\Delta G(\theta=90,B)$, obtained in films without Fe-deposition (black open circles) for 5 nm sample (a) and for 50 nm sample (b) are also plotted. As we can see, the low field magneto-conductance tip in both 5- and 50 nm $Bi_2Te_3$ samples disappears and the 2D WAL effect is completely quenched by the deposition of just one ML of Fe. This reveals that magnetic impurities can indeed destroy the $\pi$ Berry's phase of 2D topological SSs. In the language of the categorization of statistical ensembles [26], SSs without Fe deposition should be classified as belonging to the symplectic group symmetry because of the strong spin-orbit coupling which respects TRS. When time reversal invariance is violated, the ensemble goes to the unitary group symmetry which is characterized by a much smaller $B^2$ magneto-conductance [19]. The magneto-conductance of the 5 nm sample with one ML Fe deposited on it drops dramatically, and is proportional to $B^2$ (see the fitting solid line in Fig. 3 (a)). Thus our results suggest that we have observed a crossover from the symplectic to unitary classes when TRS is broken. The magneto-conductance of the 5 nm sample with 0.3 ML Fe-deposition shows the intermediate regime in this crossover process. The change from logarithmic to $B^2$ dependence also suggests that magnetic scatterings caused by randomly distributed Fe impurity spins tend to reduce $l_\varphi$. On the other hand, these topological SSs which are protected by TRS are robust against non-magnetic disorder perturbation. The $\Delta G(\theta = 90, B)$ curve obtained for the 5 nm Au-deposition sample at $T$ = 2K is shown in Fig. 3 (a) (green open triangles). The data clearly demonstrate that the magneto-conductance tip remains almost unchanged, indicating the persistence of 2D WAL effect in the $Bi_2Te_3$ sample when Au is deposited on the top interface. The very different effect of magnetic and non-magnetic impurity on our data clearly associates the 2D WAL effect we observed with the topological SSs. The disappearance of 2D WAL effect in $Bi_2Te_3$ films with Fe impurities deposited at the top film interface shows that the observed 2D WAL effect arises mainly from the top SS, in agreement with $\alpha$ =-0.39 that is obtained from the numerical fitting. The contribution to the 2D WAL effect from the bottom SS seems negligible and it might be due to the much reduced $l_\varphi$ since much more defects are expected to be present at the bottom surface due to the lattice mismatch between the $Bi_2Te_3$ and ZnSe layer.

In conclusion, we show that the observed WAL effect in $Bi_2Te_3$ thin films have contributions from both 2D surface states as well as from 3D bulk states. The 2D contributions are extracted from MR data obtained in tilted magnetic fields. The 2D WAL effect, which arises as a consequence of the $\pi$ Berry's phase, is suppressed by magnetic impurities deposited on the surface of $Bi_2Te_3$ thin film, but it is almost unaffected by non-magnetic impurities. The observed behavior of the magneto-conductance of a 5 nm $Bi_2Te_3$ film suggests our system crossovers from symplectic to unitary classes when Fe is deposited on the film surface.

We wish to acknowledge useful discussions with W. Q. Chen. This work was partially supported by the Research Grant Council of the HKSAR under Grant Numbers HKUST16/CRF/08, HKU10/CRF/08, 603407, and 604910. The PPMS facilities used for magneto transport measurements is supported by the Special Equipment Grant (SEG_CUHK06) from the UGC of the HKSAR.

**Figure Caption**

Figure 1 (Color online). (a) Temperature dependence of the resistivity for 5 nm and 50 nm $Bi_2Te_3$ films. Upper inset: Schematic illustration of the band structure of the n-type film with the Fermi level ($E_F$), the conduction band (CB), the surface states (SS), and the valence band (VB) indicated. Lower inset: Schematic drawing of the Hall bar structure and experiment geometry. (b) Normalized magneto-resistance $R/R(B=0)$ measured in a 5 nm $Bi_2Te_3$ film at various temperatures.

Figure 2 (Color online). Normalized MR of a 5 nm $Bi_2Te_3$ film (a) and a 50 nm $Bi_2Te_3$ film (c) measured in tilted $B$ fields at $T$ =2K. Solid curve in (a) is a parabolic fit to the MR data measured at $\theta$=0. Magneto-conductance as a function of normal $B$ component with the $\theta$=0 magneto-conductance subtracted for 5 nm sample (b) and 50 nm sample (d). Solid curve in (d) is a fit to $\Delta G(\theta=90, B)$ in low B-field region with Eq. (1).

Figure 3 (Color online). (a) $\Delta G(\theta=90, B)$ as a function of normal $B$ component obtained in 5 nm $Bi_2Te_3$ films at $T$ = 2K, with no metal, 0.3 ML Fe-deposition, 1 ML Fe-deposition, or 1 ML Au-deposition at the top interface. Solid curve is a parabolic fit to the $\Delta G(\theta=90, B)$ data in the presence of 1 ML Fe-deposition. (b) $\Delta G(\theta=90, B)$ as a function of normal $B$ component obtained in 50 nm $Bi_2Te_3$ films at $T$ = 2K, with no metal or 1 ML Fe-deposition at the top interface.

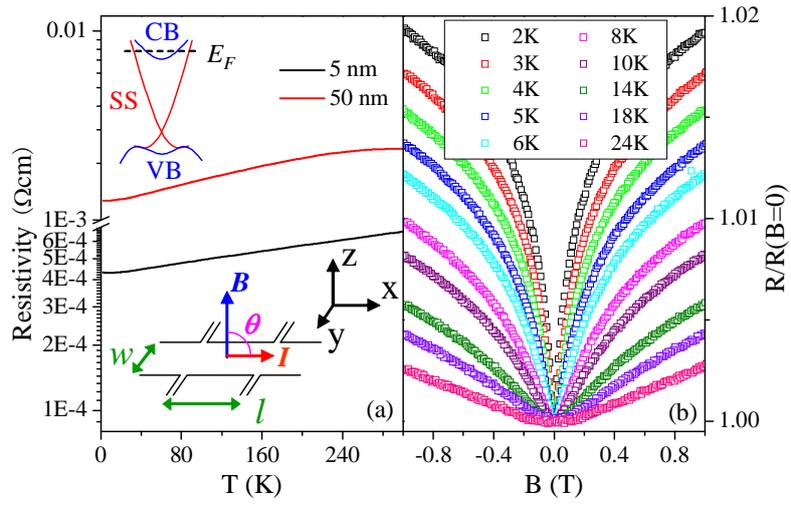

Figure 1

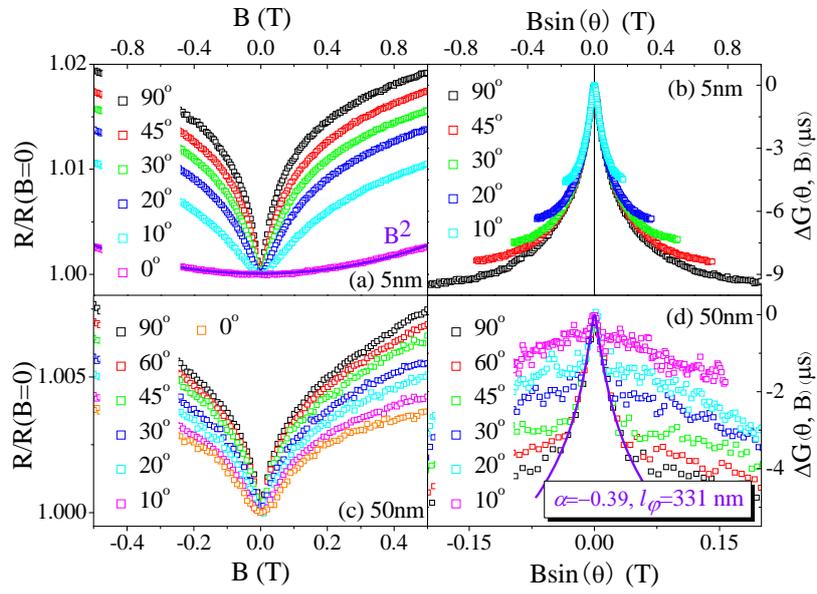

Figure 2

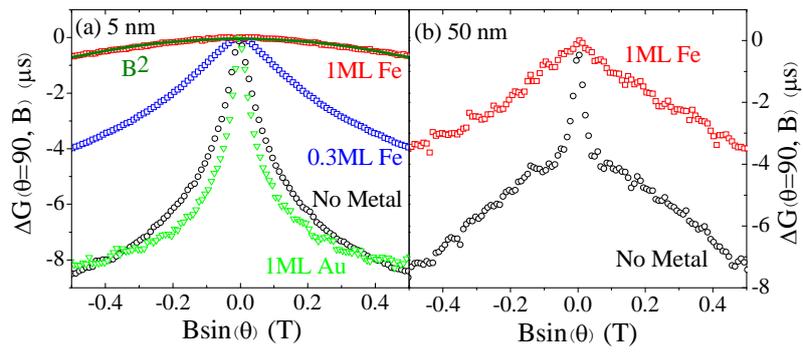

Figure 3